\documentclass[prb,preprint,showpacs,preprintnumbers,amsmath,amssymb]{revtex4}

\usepackage{graphicx}
\usepackage{dcolumn}
\usepackage{bm}
\usepackage{subfigure}

\begin{document}

\title{Diffusive and ballistic current spin-polarization in magnetron-sputtered L1$_{0}$-ordered epitaxial FePt}


\author{K. M. Seemann}
\email{phyks@leeds.ac.uk} \homepage{www.stoner.leeds.ac.uk}
\affiliation{School of Physics and Astronomy, University of Leeds,
Leeds, LS2 9JT, United Kingdom}

\author{V. Baltz}
\affiliation{School of Physics and Astronomy, University of Leeds,
Leeds, LS2 9JT, United Kingdom}

\author{M. MacKenzie}
\affiliation{Department of Physics and Astronomy, University of
Glasgow, Glasgow, G12 8QQ, United Kingdom}

\author{J. N. Chapman}
\affiliation{Department of Physics and Astronomy, University of
Glasgow, Glasgow, G12 8QQ, United Kingdom}

\author{B. J. Hickey}
\affiliation{School of Physics and Astronomy, University of Leeds,
Leeds, LS2 9JT, United Kingdom}

\author{C. H. Marrows}
\email{c.h.marrows@leeds.ac.uk}
\affiliation{School of Physics and Astronomy, University of Leeds,
Leeds, LS2 9JT, United Kingdom}


\date{\today}

\begin{abstract}
We report on the structural, magnetic, and electron transport
properties of a L1$_{0}$-ordered epitaxial iron-platinum alloy layer
fabricated by magnetron-sputtering on a MgO$($001$)$ substrate. The
film studied displayed a long range chemical order parameter of
$S\sim0.90$, and hence has a very strong perpendicular magnetic anisotropy. In
the diffusive electron transport regime, for temperatures ranging
from 2 K to 258 K, we found hysteresis in the magnetoresistance
mainly due to electron scattering from magnetic domain walls. At 2
K, we observed an overall domain wall magnetoresistance of about
0.5\%. By evaluating the spin current asymmetry $\alpha =
\sigma_{\uparrow} / \sigma_{\downarrow}$, we were able to estimate the
diffusive spin current polarization. At all temperatures ranging
from 2 K to 258 K, we found a diffusive spin current polarization of
$> 80$\%. To study the ballistic transport regime, we have performed
point-contact Andreev-reflection measurements at 4.2 K. We obtained
a value for the ballistic current spin polarization of $\sim 42$\%
(which compares very well with that of a polycrystalline thin film
of elemental Fe). We attribute the discrepancy to a difference in
the characteristic scattering times for oppositely spin-polarized
electrons, such scattering times influencing the diffusive but not
the ballistic current spin polarization.
\end{abstract}


\maketitle

\section{\label{sec:intro}Introduction}

Recent advances in the research field of current-induced
magnetization switching and current-driven magnetization
dynamics,\cite{mangin2006} as well as the developments in the hard
disk drive industry to change the magnetic storage process to
perpendicular magnetic recording cause a resurgence of interest in
ultrathin film magnetic materials with out-of-plane magnetic
anisotropy. One way to achieve this is to exploit magnetocrystalline
anisotropy in an epitaxial film.\cite{caro1998,suzuki1999,ishio2002}
This upsurge in research interest in epitaxial material exhibiting a
high perpendicular uniaxial anisotropy constant $K_\perp$ has been
stimulated especially since sputter deposition now yields epitaxial
thin films of an ordering quality comparable to molecular beam
epitaxy.\cite{schwickert2000,barmak2004,barmak2005,clavero2006} The
very high values of $K_\perp$ now available lead to domain walls of
a very narrow thickness $\delta_{\rm W}$, and this makes fundamental
physical phenomena like magnetoresistance due to electron scattering
at magnetic domain walls an easily measurable effect.
\cite{ravelosona1999,yu2000,marrows2004} After some early
experimental work,\cite{heaps1934,taylor1968,berger1978} domain wall
scattering is undergoing something of a renaissance as materials
preparation and nanofabrication technologies
improve.\cite{kent2001,marrows2005,tanigawa2006,lee2006}

This effect of an increased electric resistivity in the presence of
magnetic domain walls in a ferromagnetic thin film was measured by
Viret et al. for films of Ni and Co. \cite{viret1996} This group
developed a semiclassical model based on spin-mistracking as the
electrons cross the wall, which they used to interpret their data.
(This type of model was necessary since quantum mechanical
reflection of electrons from a domain wall potential step will be
extremely small unless the wall is of extreme
abruptness.\cite{cabrera1974a,cabrera1974b}) A more rigorous quantum
mechanical model, based on a Hamiltonian employed to calculate giant
magneto resistance in a spin-split system, was used by Levy and
Zhang to treat the same physics. \cite{levy1997} It is possible to
use this model to determine the polarization of a diffusive current
by measuring domain wall resistance.\cite{marrows2004}

This paper concentrates on the electron transport properties of the epitaxial L1$_{0}$-ordered iron alloy FePt in both the diffusive and the ballistic regime. In particular, we have measured the spin-polarization of the current in two different transport regimes: diffusive and ballistic. The strong uniaxial anisotropy, arising from the high degree of long range chemical order in our ordered alloy epilayers, leads to a dense stripe domain structure with narrow walls. These give rise to an easily measurable magnetoresistance associated with the extra resistance as the electrons pass through these walls. By using the Levy-Zhang model\cite{levy1997}, this can be used to infer the spin-polarisation of these diffusive current carrying electrons, which we find to exceed 80\% at all temperatures from 2 K to 258 K. We also determined the ballistic current spin polarization by the widely used point contact Andreev-reflection (PCAR) method.\cite{soulen1998,strijkers2001} We found a spin polarization of $\sim 42$\%, close to that reported for elemental 3$d$ ferromagnets.

\section{\label{sec:prep}Sample preparation and characterization}

\begin{figure}
\includegraphics[width=8cm]{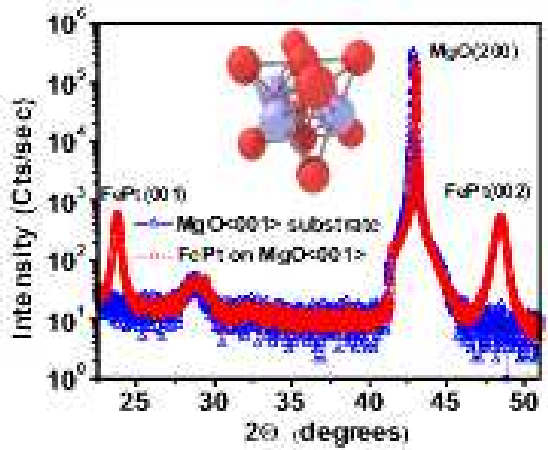}
\caption{(Color online) $\Theta$-$2\Theta$ x-ray crystallography scan
for L1$_{0}$-ordered FePt of a film thickness of 31 nm film (red
circles) The MgO$(001)$ substrate scan is included for comparison
(blue triangles). The inset illustrates an unit cell of the
face-centred tetragonal lattice of FePt. The Fe atoms (red) and Pt
atoms (blue) form alternating $a$-$b$ planes. The $c$-axis lies normal
to these planes and forms the magnetic easy axis. This is the growth
direction in our epitaxial film.}\label{figure2new}
\end{figure}

The samples were prepared by conventional dc magnetron sputter
deposition on polished MgO$(001)$ substrates. We used a
4\%-hydrogen-in-argon sputter gas mixture to prevent any film
oxidation during growth at high temperatures. The FePt magnetic thin
film  was sputtered by co-deposition directly onto the substrates at
a substrate temperature of 1000 K and at a deposition rate of
0.1-0.2 \AA/s$^{-1}$. Here we will describe the properties of a 31
nm thick film, which has one of the highest degrees of chemical
order we have achieved, having grown several dozen such samples to
optimize our deposition process.

\begin{figure}[tb]
\includegraphics[width=5cm]{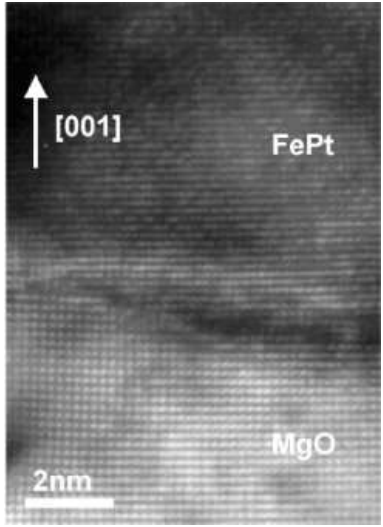}
\caption{Lattice structure as obtained from HRTEM of the
L1$_{0}$-ordered FePt thin film grown on MgO(001). The [001]
direction is the growth direction in our epitaxial film.
}\label{figure1tem}
\end{figure}

For structural characterization of the epitaxial L1$_{0}$-ordered
FePt thin films, we carried out $\Theta$-$2\Theta$ x-ray diffraction
scans using Cu-$K_{\alpha}$ radiation in order to determine the long
range order parameter $S$ $(0\leq{S}\leq{1})$ according to
\begin{equation}
S=r_{\alpha}+r_{\beta}-1=\frac{(r_{\alpha}-x_{A})}{y_{\beta}}=\frac{(r_{\beta}-x_{B})}{y_{\alpha}}.
\end{equation}
Here $x_{a}$ and $x_{b}$ are the atom fractions of the two
components, $y_{\alpha}$ and $y_{\beta}$ are the fractions of the
lattice site types $\alpha$ and $\beta$ in the ordered structure,
and $r_{\alpha}$ and $r_{\beta}$ are the fractions of each type of
lattice site occupied by the correct types of atoms, $A$ on $\alpha$
and $B$ on $\beta$.\cite{warren1969,xraytablesvolIII,xraytablesvolIV} A typical
$\theta$-$2\theta$ scan of an L1$_{0}$-ordered FePt is displayed in
Fig. \ref{figure2new}. The presence of the (001) peak is normally
forbidden by the structure factor for face-centered crystal
lattices, and so its observation here confirms that there is
preferential ordering on the alternating $\alpha$ and $\beta$
planes. The $(001)$-peak and $(002)$-peaks were fitted with
Lorentzian line shapes to yield the integrated intensities.
Following the standard procedure, described in e.g. reference
\onlinecite{warren1969}, these integrated intensities, together with
the peak positions, Lorentz polarization factors and atomic
scattering factors, can be used to give a value for $S$. We found
$S=0.90\pm0.05$ for the particular film of thickness of 31.0 nm that
we discuss in detail in this article, and routinely obtain $S>0.80$
in our sputtered films.  We calculated the film thickness via
Kiessig-fringes obtained in low-angle x-ray reflectometry
measurements.

\begin{figure}[tb]
\includegraphics[width=5cm]{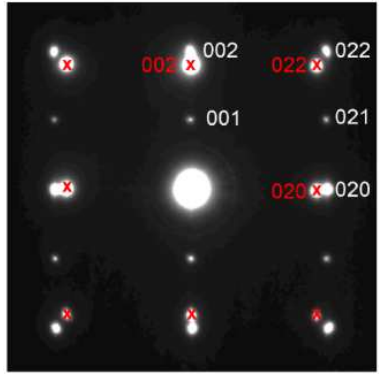}
\caption{(Color online) Selected area electron diffraction pattern
from a cross-sectional TEM sample of FePt on MgO(001). The (001)
superlattice spots confirm the L1$_{0}$-ordered FePt structure of
the FePt thin film. The diffracted spots associated with the
MgO(001) substrate are marked with red crosses.}\label{figure2tem}
\end{figure}

\begin{figure}[tb]
\includegraphics[width=7cm]{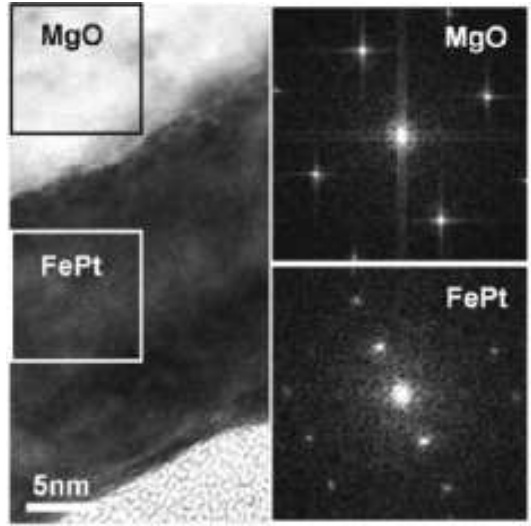}
\caption{HRTEM image of an L1$_{0}$-ordered FePt thin film grown on
MgO(001). Inset are FFT patterns obtained from the regions of the
single crystal MgO substrate and the epitaxial FePt layer marked
with boxes. It can be seen that the superlattice spots are
associated with the FePt layer. }\label{figure4tem}
\end{figure}

We performed transmission electron microscopy (TEM) on an FEI Tecnai
F20 to provide structural information on our
material.\cite{mackenzie2005} The cross-sectional high-resolution
transmission electron microscopy (HRTEM) image in Fig.
\ref{figure1tem} from a comparable sample, taken with the electron
beam aligned parallel to the $[100]$ or $[010]$ zone axis of the MgO
substrate confirms the high quality of our epitaxial FePt layers on
the MgO(001) substrates. The L1$_{0}$-ordering of the FePt was
verified by selected area electron diffraction, see Fig.
\ref{figure2tem}, which shows the diffraction pattern from a
cross-sectional TEM of the same sample of FePt on MgO(001).  The
high degree of alignment between the MgO and FePt structures is
clearly depicted. The $\{002\}$ and $\{022\}$ reflections from the
face-centered cubic MgO substrate are marked with red crosses and
show the expected four-fold symmetry associated with the $[100]$
zone axis. The remaining reflections are from the FePt layer and
index as the $[100]$ zone axis of L1$_{0}$-ordered face-centered
tetragonal FePt. The presence of the $(001)$ superlattice spots
confirms the L1$_{0}$-ordering of FePt. Using a lattice parameter of
$4.21~\rm{\AA}$ for MgO as a calibration, we obtain $a=3.85~\rm{\AA}$ and $c=3.76 \rm{\AA}$ for the FePt structure. This gives a lattice mismatch of $8.5~ \rm\%$ between the MgO substrate and FePt
layer. In Fig. \ref{figure4tem} we show another image with a larger
field of view. Fast Fourier Transform (FFT) patterns obtained from
the boxed areas in Fig. \ref{figure4tem}, and shown as insets,
confirm that the superlattice spots seen in the electron diffraction
pattern are associated with the L1$_0$ order in the FePt layer.

\begin{figure}
\includegraphics[width=8cm]{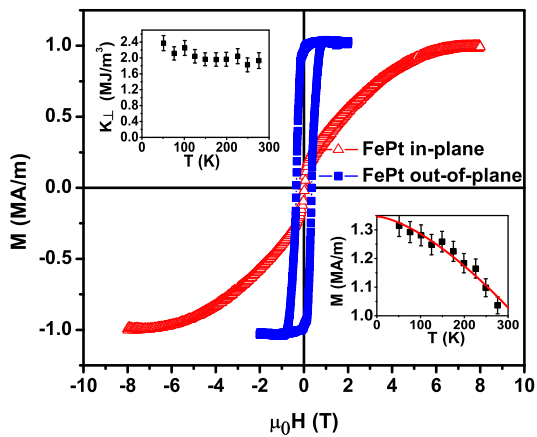}
\caption{(Color online) Hysteresis loops for the 31 nm thick
L1$_{0}$-ordered FePt film obtained by vibrating-sample-magnetometry
in the in-plane and out-of-plane geometry at $T=276$ K. The two
insets depict the uniaxial anisotropy constant $K_{\perp}$ and the
saturation magnetization, together with a Bloch-law fit, as a
function of temperature.}\label{figure3new}
\end{figure}

A quantitative analysis of the strong out-of-plane magnetic
anisotropy of the L1$_{0}$-ordered FePt film was carried out by
vibrating-sample-magnetometry (VSM) in the out-of-plane geometry as
well as the in-plane geometry, with representative hysteresis loops
shown in Fig. \ref{figure3new}. The uniaxial magnetocrystalline
anisotropy constant $K_{\perp}$ was calculated
from\cite{chikazumi1998}
\begin{equation}
K_{\perp}=\mu_{0} \int_{0}^{M_{\rm sat}}(H_{\rm hard-axis}-H_{\rm
easy-axis})dM + K_{\rm demag},
\end{equation}
where the extra term $K_{\rm demag}=\frac{1}{2}\mu_0M_{\rm sat}^2$
accounts for the demagnetization field within the sample. $H_{\rm hard-axis}$ and $H_{\rm easy-axis}$ are the magnetic fields applied in and normal to the film plane respectively. For $T=276$
K, we found $K_{\perp}=1.9\pm0.2$ MJm$^{-3}$, $M_{\rm
sat}=1.0\pm0.1$ MAm$^{-1}$, and $A=14.2\pm4$ fJm$^{-1}$. We deduced
the zero-Kelvin exchange stiffness $A$ from a
$T^{\frac{3}{2}}$-Bloch law fit of the temperature dependence of the
saturation magnetization\cite{chikazumi1998} and assumed that $A(T)$
follows a mean-field behavior $\propto M_{\rm sat}(T)$: the data and
fitted curve are shown in the lower right inset of Fig.
\ref{figure3new}. The temperature dependence of $K_{\perp}$ is shown
in the upper left inset. Experimental micromagnetic data was not
available for $T<50$ K due to the large signal that arises at low temperatures caused by paramagnetic impurities in the
substrate (typically at the parts per million level in epi-ready MgO). The values we obtain compare reasonably well with the micromagnetic parameters recently reported for L1$_{0}$-ordered FePt
thin films grown by molecular beam epitaxy and magnetron-sputtering
of other groups. \cite{farrow1996,daalderop1991,barmak2005,clavero2006,inoue2006,okamoto2002}.

\begin{figure}
\includegraphics[width=4cm]{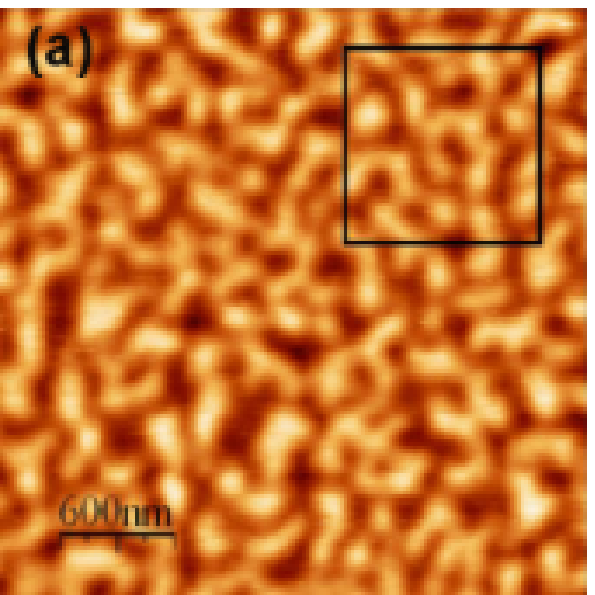}
\includegraphics[width=4cm]{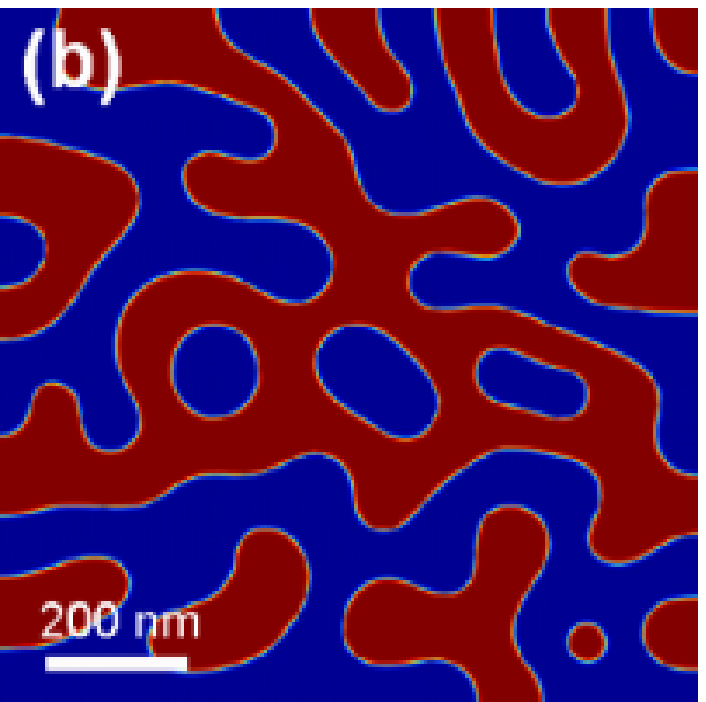}
\caption{(Color online) Magnetic force image (a) showing the typical
labyrinth domain structure of the demagnetized L1$_{0}$-ordered FePt
film in zero magnetic field. The bright and dark areas mark magnetic
domains of opposite perpendicular magnetization. A power spectrum
analysis of such images leads to an average domain width of
$\sim170\pm20$ nm. Micromagnetic simulation (b) of the domain
structure of a $1\mu m\times 1\mu m$ FePt thin film of 30 nm
thickness using the OOMMF code and experimentally determined
micromagnetic parameters. The red and blue regions mark areas of
opposite perpendicular magnetisation. The typical domain width can
be seen to be $D \sim 150$ nm. The black square in panel (a) shows a region of the
same size as that simulated and shown in
panel (b).}\label{figure4anew}\label{figure4bnew}
\end{figure}

We imaged the magnetic domain structure of FePt by magnetic force
microscopy (MFM) at room temperature in zero field, as shown in Fig.
\ref{figure4anew} (a). The sample was demagnetized using an
alternating magnetic field of decreasing amplitude. The cantilevers
had a resonant frequency of 65 kHz and a spring constant of 1-5 N/m.
The CoCr-coated Si tip was vertically magnetized prior to imaging.
For optimal contrast we kept the tip-surface distance constant at a
value in the range 20-25 nm. The average magnetic domain width of
the demagnetized state was obtained by a power-spectrum analysis and
resulted in a domain width of $D=(170\pm15)$nm for the 31.0 nm thick
sample. The domain structure exhibits the typical interconnected
dense stripe domain structure known from L1$_{0}$-ordered binary
iron alloys.\cite{thiele1998,barmak2005,gehanno1997}

We carried out micromagnetic simulations of this domain structure in the sample using the
\textsc{oommf} code,\cite{nist} the results of which are shown in Fig. \ref{figure4bnew} (b). The cell size
used was $(1\times1)$ $\rm{nm}^2$ within the film plane and 15 nm
perpendicular to the film plane, and a six-nearest-neighbor
exchange interaction for the magnetic energy terms of adjacent cells
was employed. Although thermal activation effects are not taken into
account in this type of micromagnetic code, we were nevertheless
able to simulate the domain structures in our material at finite
temperatures using the appropriate values of the micromagnetic
parameters $A(T)$, $K(T)$ and $M(T)$ as determined from
vibrating-sample-magnetometry, as we are not concerned with thermal activation effects when determining the equilibrium domain structure. Our simulation yields an average
domain width $D \sim 150$ nm at room temperature for a 30 nm thick film obtained from a
Fourier analysis of the \textsc{oommf} output. The analytical result obtained from the Kaplan-Gehring model\cite{kaplanandgehring} is $\sim 130$ nm at room temperature and $\sim 90$ nm at 50 K, so the temperature dependence of the domain strip width is quite fairly weak.

We also estimated the average width of an individual domain wall analytically from
\begin{equation}
\delta_{\rm W}\simeq \pi\sqrt{\left(\frac{A}{K_{\perp}}\right)}.
\label{wallwidth}
\end{equation}
Using our experimentally determined micromagnetic parameters, we
obtain $\delta_{\rm W} \sim$ 8-9 nm at all temperatures
in good agreement with our micromagnetic simulations. Such narrow wall thicknesses are extremely difficult to measure experimentally. Such narrow Bloch-type domain walls may be found in many hard magnets materials such as NdFeB or SmCo. Their effect on thin film electron transport properties such as domain wall resistance is particularly
interesting. Narrower walls are available only in a very few magnetic materials (e.g. at low temperatures in SrRuO$_3$ \cite{klein2000}).



\section{\label{sec:transport}Electron transport properties}

\subsection{\label{sec:diffusive}Transport in the diffusive regime}
\begin{figure}
\includegraphics[width=8cm]{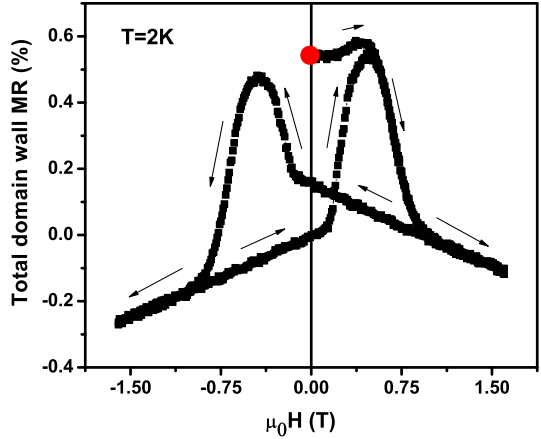}
\caption{(Color online) Magnetoresistance vs. applied field of
L1$_{0}$-ordered FePt obtained by a DC in-line 4-terminal
measurement at $T=2$ K in the perpendicular field geometry.
 The red dot marks the resistance in the demagnetized state of the
sample, at the start of the virgin branch of the hysteresis
loop.}\label{figure5new}
\end{figure}

We first describe the diffusive transport properties of
L1$_{0}$-ordered FePt. We performed magnetotransport measurements at
temperatures ranging from 2 K to 258 K using an in-line 4-terminal
set-up with the magnetic field applied normal to the film plane. We
found a hysteretic part of the magnetoresistance of $MR_{\rm
Domain}=0.55\%$ at 2 K and $MR_{\rm Domain}=0.26\%$ at 258 K,
associated with the creation and annihilation of domain walls as the
film switches its magnetization direction. A typical MR hysteresis
loop is shown in Fig. \ref{figure5new}. From previous studies on
L1$_{0}$-ordered FePd thin films,\cite{marrows2004} we know that a
sufficiently high quality factor $Q = 2K_{\perp}/\mu_0 M^2$ as similarly in this case, $Q\approx2.2$, 
is a good indication that the anisotropy magnetoresistance (AMR)
contribution of N\'{e}el closure caps on the domain walls cannot
account for this effect and is small enough to be neglected. The
asymmetry of the MR loops arises through the extraordinary Hall
effect, caused by large spin-orbit interaction in FePt, and the
minute misalignment of our voltage probes. This effect can be easily
subtracted to give the true domain wall MR. We will discuss the
extraordinary Hall effect in films such as these in more detail
elsewhere.

\begin{figure}
\includegraphics[width=8cm]{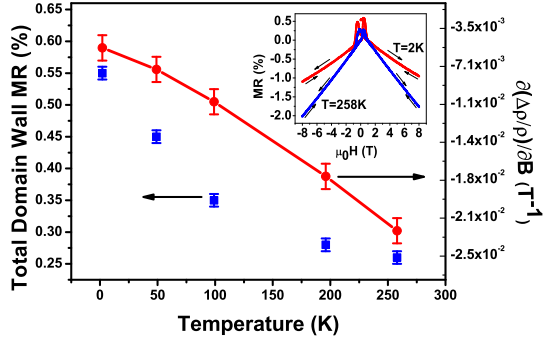}
\caption{(Color online) Total domain wall MR (blue squares) and high
field MR slope (solid red circles) vs. temperature for a L1$_{0}$-ordered
FePt thin film of a thickness of 31 nm. The solid line is a fit to
the data as described in the text. The inset shows hysteretic MR
loops obtained for L1$_{0}$-ordered FePt at $T=2$ K and $T=258$ K in
the perpendicular geometry. A strong linear high field
magnetoresistance is evident in both cases.}\label{figure6new}
\end{figure}

We find the domain wall MR to be approximately twice as large as
compared to those reported on L1$_{0}$-ordered FePt films grown by
molecular beam epitaxy,\cite{yu2000} even though our film had a
rather high electrical resistivity of $\rho=35 \mu\Omega$ cm (at 2
K) with at a residual resistivity ratio of $RRR=2.4$. The
temperature dependence of the domain wall MR is shown in Fig.
\ref{figure6new} and exhibits an almost linear behavior with
temperature. We observed a nearly complete suppression of any K\"{o}hler
magnetoresistance ($\propto B^2$) at high magnetic fields and low
temperatures (Fig. \ref{figure6new}), due to the high film
resistivity giving rise to a very small value of $\omega_{\rm C}
\tau$. Hence, we attribute the much weaker effect of K\"{o}hler MR, as compared to FePd,\cite{marrows2004} to considerably
reduced electron mean free paths.

In addition to the hysteretic part of the MR we observe a reversible
linear part at high fields.  We have extracted the high field MR
slopes $\partial(\Delta \rho/ \rho) / \partial B$ at an applied
magnetic field of 5 T at various temperatures (Fig.
\ref{figure6new}). In Fe, Co, and Ni, such a negative and linear MR
was found by Raquet et al.\cite{raquet2001} to be caused the
influence of a magnetic field on the spin mixing resistivity. There,
the main role is played by spin-flip $s$-$d$ inter-band and
intra-band scattering due to electron-magnon scattering. The data in
Fig. \ref{figure6new} can be fitted with the expression given by
Raquet et al. \cite{raquetprb2002} quite well, with the exception
that a substantial linear MR ($\partial(\Delta \rho/ \rho) /
\partial B = -0.0054$ T$^{-1}$) remains even at the lowest
temperatures in this film, which must be added as an additional
constant term. Subsequently, we could fit the temperature dependence
of the high-field MR slope (Fig. \ref{figure6new}) according to
Raquet et al.\cite{raquetprb2002}. This procedure yields a magnon
mass renormalization constant of $d_1\approx-6.0\times 10^{-7}$,
which is comparable with that of pure 3$d$
metals,\cite{stringfellow1968}, and less negative than that found
previously in MBE-grown FePd.\cite{marrows2004} We do not have a
simple explanation for the temperature independent part of the
high-field linear MR, although we note that magnetoresistances in
thin films can take on a variety of unexpected forms.\cite{pippard}

Furthermore, we have used our magnetoresistance data to compute the
spin-current asymmetry parameter $\alpha$ in L1$_{0}$-ordered FePt
based on the Levy-Zhang spin-mistracking model.\cite{levy1997} The
spin asymmetry of the current depends on the spin-resolved conductivities
$\sigma_{\uparrow}$, $\sigma_{\downarrow}$ (or spin-resolved
resistivities $\rho_{\uparrow}$, $\rho_{\downarrow}$) of the majority
and minority spin channels and is given by
$\alpha=\sigma_{\uparrow}/\sigma_{\downarrow}=\rho_{\downarrow}/\rho_{\uparrow}.$

\begin{figure}
\includegraphics[width=8cm]{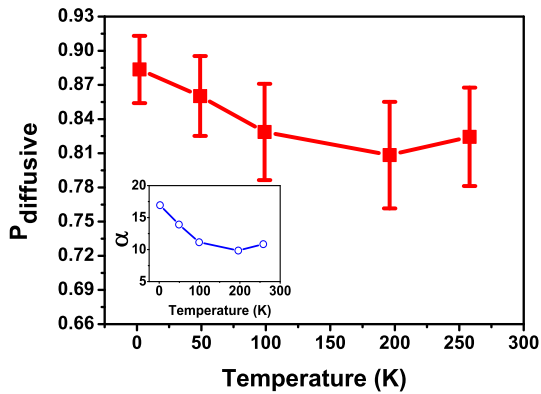}
\caption{(Color online) The $T$ dependence of the diffusive current
spin polarisation $P_{\rm diffusive}$ and in the inset the spin
resistivity asymmetry $\alpha$.}\label{figure8new}
\end{figure}

The Levy-Zhang model describes the MR only in the wall region,
whereas we have measured our entire film. We estimated the volume fraction of
walls by measuring the total wall length $\lambda_{\rm W}$
in the MFM image of scan width $\Lambda$ and multiplying this by the
wall thickness $\delta_{\rm W}$ to obtain the total area occupied by of walls, out of a total area of $\Lambda^2$.
This procedure yields a volume fraction accounting for the fact that
we do not have a parallel stripe domain state, but rather a
labyrinth structure, and yields a value approximately 1.3 times greater than the ideal stripe domain value $\delta_{\rm W}/D$. Thus one obtains for an isotropic labyrinth
domain state a domain wall magneto-resistance of\cite{marrows2004}
\begin{equation}
\frac{\Delta\rho}{\rho}=\frac{1}{5}\left( \frac{\lambda_{\rm{W}}
 \delta_{\rm W}}{\Lambda^2}\right) \left(\frac{\pi\hbar^2
k_{\rm F}}{4mJ\delta_{\rm
W}}\right)^2\frac{(\alpha-1)^2}{2\alpha}\left(4+\frac{10\sqrt{\alpha}}{\alpha+1}\right),
\label{LZmodel}
\end{equation}
where $k_{\rm F}$ is the Fermi wavevector, $m$ is the effective electron
mass and $J$ is the Stoner exchange-splitting energy. It is to be
noted that this formula yields the same $\Delta\rho/\rho$ for both
$\alpha$ and $1/\alpha$, equivalent to saying that we are
insensitive to the sign of the polarization. Based on the assumption
that the majority carriers are $s$-like, we take $m$ to be equal to
the free electron mass, assume the value of $k_{\rm F}$ to be 2
\AA$^{-1}$ (a typical value for a metal), and take a value for the
Stoner exchange splitting to be $J(0)=2.0$ eV based on the splitting
of the density of states seen in the results of band structure
calculations.\cite{maclaren2005} We appropriately scaled the domain wall volume fraction based on the analytical values for $\delta_{\rm W}$ and $D$ for different temperatures. The inset of Fig. \ref{figure8new} shows the
spin current asymmetry $\alpha$ of L1$_{0}$-ordered FePt calculated according
to Eq. \ref{LZmodel}. A strong temperature dependence of $\alpha$ is
clearly visible, with a decay of the spin current asymmetry from
$\alpha=16$ to $\alpha=10$ in the temperature range between 2 K and
258 K. It is then straightforward to obtain the diffusive current
spin polarization of L1$_{0}$-ordered FePt from
\begin{equation}
P_{\rm diffusive}=\left(\frac{\sigma_{\uparrow} -
\sigma_{\downarrow}}{\sigma_{\uparrow}
+\sigma_{\downarrow}}\right)=\left(\frac{\alpha -1}{\alpha +1}
\right),
\end{equation}
which is shown together with the spin current asymmetry in Fig.
\ref{figure8new}. For a temperature of 258 K, we found a diffusive
spin current polarization of $P_{\rm diffusive}=0.82\pm0.04$,
whereas $P_{\rm diffusive}=0.88\pm0.02$ at $T=4.2$ K. The
uncertainties in these polarization values are determined from the
uncertainties in the $\alpha$-values, which in turn were computed
using Eq. \ref{LZmodel}, and also taking into account the
uncertainties of the micromagnetic parameters $K_\perp$, $M$ and $A$
taken to evaluate the domain wall dilution. The final uncertainties are small since the diffusive polarization is rather insensitive to the value of $\alpha$ when $\alpha \gg 1$.

\subsection{\label{sec:ballistic}Transport in the ballistic regime}

We performed point contact Andreev reflection (PCAR) measurements in
order to directly probe for the ballistic current spin polarization
of L1$_{0}$-ordered
FePt.\cite{jong1995,soulen1998,upadhyay1998,strijkers2001,nadgorny2001,ji2001,marrows2005}
The concept of this method is based on the fact that for applied
bias voltages within the gap of the superconductor, it is physically
impossible to inject or extract single electrons, but only Cooper
pairs. As the Andreev reflection process\cite{andreev1988} is the
coherent back reflection of a charge carrier hole into the
ferromagnetic sample following the capturing of an opposite spin
electron to form a Cooper pair inside the superconducting tip, one
essentially probes for the number of unpaired electrons,
straightforwardly giving the ballistic spin current polarization of
the ferromagnet.

\begin{figure}
\includegraphics[width=8cm]{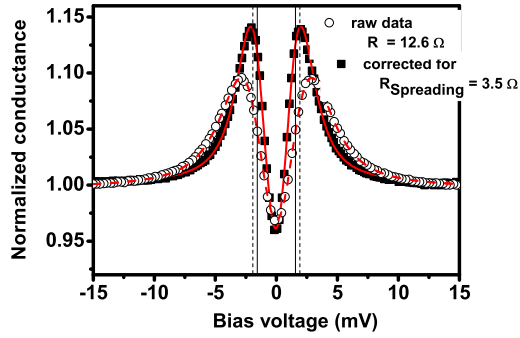}
\caption{(Color online) Normalized conductance vs. bias voltage as
obtained by point contact Andreev reflection (PCAR) at $T=4.2$K
before (circles) and after (squares) correction from the spreading
resistance as defined in the text, together with the respective fits
according to the modified BTK model (dashed and plain red lines).
The fitting parameters are discussed in the text and given in the
following figure. The vertical lines indicate the values of the
deduced gaps before (dashed line) and after (plain line)
correction.}\label{figure9new}
\end{figure}

The point contact was controlled mechanically at 4.2K, in a liquid
helium bath, between a superconducting niobium tip and the FePt thin
film. We used the same sample as for the diffusive transport
characterization. A bias voltage was applied across the point
contact and the differential conductance was recorded via a
four-probe technique. AC lock-in detection with a $0.1$ mV amplitude
and a 5 kHz frequency was used. The tips were repeatedly brought
into contact with the sample and the dependences of the differential
conductance with the sample-tip bias voltage were recorded for
various contact resistances. A typical curve is shown in Fig.
\ref{figure9new}. As also displayed in Fig. \ref{figure9new}, the
data were corrected from the contribution of spreading resistances
within the film, as deduced from our measurements. Spreading
resistances are commonly found when the resistances of the point
contact (of around 10 $\Omega$ in our case) are of the same order of
magnitude as the resistance of the film (here around 80 $\Omega$),
and a common tell-tale sign is that the superconducting gap is
significantly overestimated. It is then necessary to correct both
voltage bias and differential conductance data for this additional
series resistance.\cite{woods2004} However, the effect of correction
on polarisation is not large in our case, since the ratio of sub-gap
to quasiparticle conductance never strays too far from unity at any
value of bias, due to the polarisation is being close to 50 \%.

\begin{figure}
\includegraphics[width=8cm]{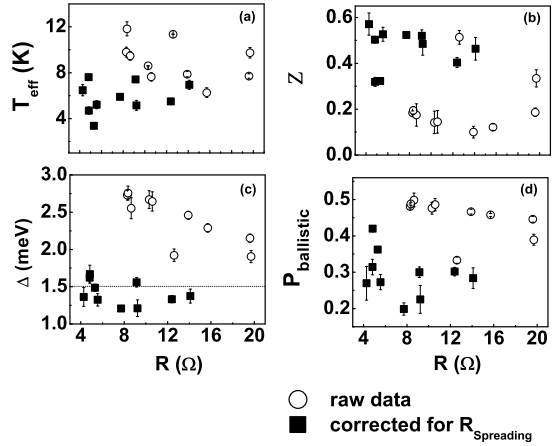}
\caption{Dependance of the fitting parameters employed in the
modified BTK model, and described in the text, with the point
contact resistance.}\label{figure85}
\end{figure}

From the typical resistances of the point contacts ranging between 4
and 15 $\Omega$, and using the Sharvin formula, we calculate an
efficient point contact characteristic size of around 5 to 15
nm.\cite{bugoslavsky2005,woods2004} Such a value is much smaller
than the characteristic micron-size of the terminated apex of our
tip, as measured by scanning electron microscopy. Indeed, as it is
usually the case, our contact results in multiple efficient
nanometric point contacts,\cite{bugoslavsky2005} where electron
transport across the ferromagnet-superconductor interface is
ballistic.

The conductance vs. bias voltage data were fitted in the standard
way, employing a modified Blonder-Tinkham-Klapwijk (BTK) model
\cite{blonder1982}, which describes the crossover from metallic to
tunnel junction behavior of a microconstriction contact between a
semiconductor and a superconductor based on the Bogoliubov
equations. Four numerical fitting
parameters\cite{strijkers2001,bugoslavsky2005} are employed to fit
the measured conductance curves and thus determine the bulk current
spin polarization of the sample: the effective temperature, $T_{\rm{
eff}}$; the barrier strength, $Z$, which accounts for the cleanness
of the interface (e.g. an infinite $Z$ accounts for a tunnel
transport regime); the superconducting gap, $\Delta$ ($\sim1.5$ meV
for elemental bulk niobium); and the spin polarization $P_{\rm
{ballistic}}$. The dependences of the fitting parameters on the
point contact resistance ($R$) are plotted in Fig. \ref{figure85},
among with the resulting fits of the raw data, for comparison.

From the data shown in Fig. \ref{figure85}(a), it can be seen that on average
$T_{\rm{eff}}$ is larger than the 4.2 K real temperature of the
experiment.\cite{strijkers2001,kant2002} Such differences between
effective and real temperatures have already been reported and are
beyond the scope of this article. They are ascribed to weaknesses in
the model since $T_{\rm{eff}}$ not only accounts for thermal
activation but also includes other effects that result in a
broadening of the Fermi-Dirac function such as the electron Fermi
velocity mismatch between the tip and the sample or the presence of
a thin remaining oxide layer at the surface. Moreover, this can also
represent any spread in the properties of different parallel
nanocontacts formed by the tip and sample. To avoid confusion,
$T_{\rm{eff}}$ is sometimes referred to, in the literature, as a
broadening factor.

\begin{figure}
\includegraphics[width=8cm]{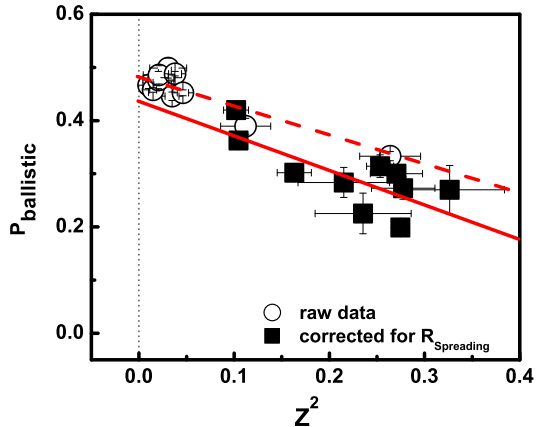}
\caption{(Color online) Ballistic current spin polarization vs.
square of the superconductor-ferromagnet interface transparency
parameter $Z$ for L1$_{0}$-ordered FePt. The extrapolation of the
least-squares fit (lines) onto the ordinate gives the bulk spin
polarization of the current.}\label{figure10new}
\end{figure}

From the data shown in Fig. \ref{figure85}(b), it can be seen that for a given tip, there
is no clear correlation between the point contact resistance and
$Z$. It had been ascribed to the fact that for contacts of the same
nature, $R$ is mainly determined by the size of the contact rather
than its cleanness.\cite{strijkers2001} As observed in Fig.
\ref{figure85}(c), the values of the tips superconducting gaps are
in agreement with those of the bulk Nb. Note that the initial large
values of the superconducting gap as deduced from fits of the raw
data are indeed the signature of spreading resistances. Figure
\ref{figure85}(d) shows that the spin polarization does not depend
on $R$ in an easily observable way. Rather note that the fitted spin
polarization seems to systematically depend on $Z$. Here we find an
acceptable agreement with a quadratic reduction in
$P_{\rm{ballistic}}$ with $Z$,\cite{strijkers2001,ji2001} as shown
in Fig. \ref{figure10new}. The relevant value of the spin
polarization is known to be the one extrapolated in the case of a
perfectly transparent interface (i.\ e.\ when $Z=0$). We find a
ballistic spin polarization of $P_{\rm{ballistic}}=(0.42 \pm 0.05)$
for our FePt film.

We note that we obtain similar values of the polarization when we do
not apply any correction for the spreading resistance in our FePt
film.\cite{woods2004} This value is moreover close to that reported
for elemental iron using the same technique, i.e. $P_{\rm
ballistic}=0.46\pm 0.03$.\cite{kant2002} We also note that we
obtained the same value, to well within the error bar, when
analyzing data taken on the sample in the remanent state or in the
demagnetized state (not shown). It is actually not surprising, as
this method is sensitive only to the magnitude and not the direction
of the spin polarization. It is however important to notice that
this shows that in this case the effect of any stray fields due to
the domain structure on the superconductivity of the tip is
negligible.

\section{\label{sec:conc}Discussion and Conclusion}

We have determined both the diffusive and ballistic transport
spin-polarization in high quality epitaxial sputtered L1$_{0}$ FePt
thin films. In the diffusive electron transport regime, we used
magnetoresistance of domain walls along with a modified form of the
Levy-Zhang model to determine the spin current asymmetry and hence
the diffusive spin polarization of a dc current flowing in L1$_{0}$
FePt. In the ballistic electron transport regime, we extracted the
spin polarization directly from point contact Andreev reflection
measurements at 4.2K.

Comparing the polarization in the ballistic transport regime to that
in the diffusive, we find that at liquid He temperatures, where the
comparison is direct, $P_{\rm diffusive}$ is substantially higher.
In fact, to change the value of $P_{\rm diffusive}$ to be equal to
that measured for $P_{\rm ballistic}$ by PCAR, it is necessary to
change $\alpha$ by a factor of a little over 24. This is because the
diffusive polarisation is extremely insensitive to $\alpha$ when it
is large. Whilst the exact value can be modified by making a
different choice for the value of $k_{\rm F}$ or $m$, it is not
possible to get a value of $P_{\rm diffusive}$ that is close to
$P_{\rm ballistic}$ with a physically reasonable set of parameters.
To do so, it is necessary to choose a value for $m$ that is less
than the free electron mass, extremely unrealistic for a transition
metal alloy. With regards to $k_{\rm F}$, we chose $2~ \rm{\AA}^{-1}$
as a representative value of $k_{\rm F}$ for a metal. In order to
obtain a value of $P_{\rm diffusive}$ to match the PCAR value of , we would need $k_{\rm F}=6.9 \rm{\AA}^{-1}$,
corresponding to an electron density of about $10^{31} \rm{m^{-3}}$,
three orders of magnitude too high for a metal, and placing the
Fermi surface in the 4th Brillouin zone.

One would not expect that $P_{\rm diffusive}$ and $P_{\rm
ballistic}$ should be the same in any case. It was pointed out by
Mazin \cite{mazin1999} that for a fairly transparent ballistic
contact, the conductivity for given spin sub-band $\propto g(E_{\rm
F}) v_{\rm F}$, whilst for ordinary diffusive transport the
conductivity $\propto g(E_{\rm F}) v_{\rm F}^2 \tau$. It is on these
parameters that the spin polarization depends:\cite{mazin1999}
compare equation \ref{Pball} with equation \ref{Pdiff}. The
ballistic current polarization is given by

\begin{equation}
P_{\rm{ballistic}}=\frac{g_{\uparrow}(E_F)\nu_{F,\uparrow}-g_{\downarrow}(E_F)\nu_{F,\downarrow}}{g_{\uparrow}(E_F)\nu_{F,\uparrow}+g_{\downarrow}(E_F)\nu_{F,\downarrow}}.
\label{Pball}
\end{equation}

On the other hand, taking into account spin-dependent electron
scattering events within the Drude theory, the diffusive current
polarization is given by

\begin{equation}
P_{\rm{diffusive}}=\frac{g_{\uparrow}(E_F)\nu^2_{F,\uparrow}\tau_\uparrow-g_{\downarrow}(E_F)\nu^2_{F,\downarrow}\tau_\downarrow}{g_{\uparrow}(E_F)\nu^2_{F,\uparrow}\tau_\uparrow+g_{\downarrow}(E_F)\nu^2_{F,\downarrow}\tau_\downarrow},
\label{Pdiff}
\end{equation}
which involves a spin-dependent relaxation time $\tau$, besides the
band structure parameters like the density of states $g(E_F)$ and
the square of the Fermi velocity $\nu_F$. Our work could act as a
stimulus for detailed band-structure calculations needed to average
$v_{\rm F}$ and $v_{\rm F}^2$ over the whole Fermi surface in order
to make quantitative comparisons, but we would like to note that the
scattering rate $1/\tau$ is seen to be the decisive parameter here.
It is not unreasonable to expect that $\tau_\uparrow \neq
\tau_\downarrow$ in a ferromagnet such as this, where scattering
from defects and impurities occurs at different rates for carriers
of different spin.\cite{ohandley} Our parameters for scattering
within the FePt metal are within the range of those reported for
various impurities introduced as scattering centres into a 3d
magnetic matrix. \cite{dorleijn1976,campbellandfert,vandenberg}
Moreover, the parameter $\beta$ that appears in drift-diffusion
models of the current perpendicular to plane giant magnetoresistance
plays the role of the spin-polarization of the diffusive
conductivity within the bulk of a magnetic layer. Values for $\beta$
of up to 0.9 have been found for some commonplace 3d ferromagnetic
alloys.\cite{bass1999} We therefore explain the much higher values
of $P_{\rm{diffusive}}$ as compared to $P_{\rm{ballisitic}}$ as
arising from the asymmetry in the scattering rates for spin-up and
spin-down for scattering from vacancies, impurities, and anti-site
defects in the L1$_{0}$ structure, which lead to additional
polarization in the diffusive current over and above that from the
band structure alone. Meanwhile only the electronic structure
affects the polarization obtained from the PCAR method. It is worth
noting that even an unpolarized electron gas can carry a diffusive
current of finite spin-polarization in the presence of
spin-dependent relaxation times. Hence an appropriate ratio of
spin-dependent scattering rates can considerably amplify (or, in
unfavorable circumstances, attenuate) an intrinsic
spin-polarization in terms of number density when a current starts
to flow. \\This is significant, since it is $P_{\rm{diffusive}}$
that is the relevant parameter entering into theories of current
driven wall
motion\cite{tatara2004,zhang2004,li2004,thiaville2004,thiaville2005,tserkovnyak2006,vanhaverbeke2007,krueger2007}
indeed it was in this context that Berger suggested, almost thirty
years ago, that a measurement of the pressure exerted on a wall in
ferromagnet could be used to determine current
polarisation\cite{berger1978}. We expect that the results of this
measurement of $P_{\rm{diffusive}}$ should be in accord with that
measured by domain wall resistance. However, due to the dearth of
experimental data for diffusive values, when experimental data are
interpreted in terms of these theories, lower value of polarization
determined from a non-diffusive transport regime such as PCAR is
often the only available one to
use\cite{beach2006,thomas2006,hayashi2007,meier2007}. However, we
can see from the results in this article that these values
significantly underestimate the real degree of spin-polarization,
and hence the rate at which spin angular momentum is transferred to
a domain wall. We also anticipate that the diffusive current
polarization in more technologically relevant materials such as
Permalloy might be measured by making suitable nanostructures that
exploit geometrical confinement to form narrow domain
walls\cite{bruno1999} in order to yield a sufficiently large domain
wall resistance to be easily measured.

\begin{acknowledgments}
The authors would like to thank A. T. Hindmarch, G. Burnell, M. Ali
and M. C. Hickey for fruitful discussions. The technical assistance
of J. Turton, M. Patel, L. Harris, A. Price, and B. Miller is also
gratefully acknowledged. This work was supported by the UK EPSRC
through the Spin@RT research programme. VB acknowledges the
financial support provided through the European Community's Marie
Curie actions (Research Training Networks) under contract
MRTN-CT-2003-504462, ULTRASMOOTH.
\end{acknowledgments}


\end{document}